# Tailoring and enhancing spontaneous two-photon emission processes using resonant plasmonic nanostructures

Alexander N. Poddubny[1,3], Pavel Ginzburg[2*], Pavel A. Belov[1,4], Anatoly V. Zayats[2], and Yuri S. Kivshar[1,5]

[1]National University for Information Technology, Mechanics and Optics (ITMO), St. Petersburg 197101, Russia

[2]Department of Physics, King's College London, Strand, London WC2R 2LS, United Kingdom

[3]Ioffe Physical-Technical Institute of the Russian Academy of Sciences, St. Petersburg 194021, Russia

[4]School of Electronic Engineering and Computer Science, Queen Mary University of London, London E1 4NS, United Kingdom

[5]Nonlinear Physics Center and Center for Ultrahigh-bandwidth Devices for Optical Systems (CUDOS), Australian National University, Canberra ACT 0200, Australia

Abstract: The rate of spontaneous emission is known to depend on the environment of a light source, and the enhancement of *one-photon* emission in a resonant cavity is known as the Purcell effect. Here we develop a theory of spontaneous *two-photon* emission for a general electromagnetic environment including inhomogeneous dispersive and absorptive media. This theory is used to evaluate the two-photon Purcell enhancement in the vicinity of metallic nanoparticles and it is demonstrated that the surface plasmon resonances supported by these particles can enhance the emission rate by more than two orders of magnitude. The control over two-photon Purcell enhancement given by tailored nanostructured environments could provide an emitter with any desired spectral response and may serve as an ultimate route for designing light sources with novel properties.

* Corresponding author: pavel.ginzburg@kcl.ac.uk

Spontaneous two-photon emission (STPE) is a second-order quantum process where an excited electron decays to its ground state by the simultaneous emission of a photon pair. The phenomenon allows any combination of photon energies satisfying total energy conservation, resulting in a very broad emission spectrum. Multi-photon states, and in particular those involving just two, are playing key roles in quantum cryptography and computing. For example, different types of quantum sources, such as entangled state generators [1], amplifiers [2] and lasers [3] may exhibit unusual properties, such as two-photon coherent states [4], quantum noise quenching and squeezing [5], and continuous variable generators for quantum cryptography [6]. STPE is one of the fundamental quantum phenomena, and its applications span from information and communication technologies to biological imaging and related research [7].

STPE has recently been observed in optically pumped and electrically driven semiconductor structures [8]. STPE from semiconductors is competitive with other nonlinear processes, such as spontaneous parametric down-conversion (SPDC), and it may become a new route for quantum devices based on nonclassical states of light [9]. Efficient, high-rate, compact, room-temperature sources of entangled photon pairs, based on STPE have also recently been proposed [10].

In this Letter we develop the general theoretical approach for the STPE problem for a point-like emitter embedded within an arbitrary photonic environment, emphasising the proper treatment of dispersion and absorption of the material components. As an example we considered spherical metallic nanoparticles and demonstrate by tuning the parameters of this nanoplasmonic system that it exhibits control over the emission spectrum and strong Purcell-like enhancement of the radiative decay of general two-photon emitter.

Since the pioneering work of Purcell [11], it is well known that the strength of light-matter interaction can be significantly influenced by engineering the local structure of electromagnetic modes, e.g., in a cavity. When such manipulation of light is required in the optical spectral region a number of different systems may be used, such as photonic crystals [12], metal nanostructures [13,14], and metamaterials [15,16,17]. Thus far, the majority of these systems have only been applied to *one-photon processes*. The Purcell factor ($P_n$ - *n* photons involved) of high-order processes is of great importance, since they are initially weak and may be improved as roughly the n$^{th}$ power of the one-photon ($P_1$) enhancement. The STPE from a bulk semiconductor layer interfaced with a metallic nano-antenna was recently demonstrated to exhibit up to 10$^3$-fold overall enhancement [18]. The proper theoretical treatment is essential for these processes where photonic states may have some level of non-classicality.

The quantization of electromagnetic field in the presence of material bodies requires a significant care. Material dispersion and absorption (related to each other by Kramers–Kronig relations) increase the complexity of a canonical Hamiltonian quantization, since additional 'material' degrees of freedom must be included. The easiest approach for field quantization, based on field expansion over the classical electromagnetic modes of a structure and known as mode decomposition, is not applicable to these lossy systems [19]. One of the rigorous techniques for field quantization is based on an introduction of local noise operators, and it results in a point-by-point quantization of an electromagnetic domain. Here, the field operators are strictly related to the classical electromagnetic Green's functions [20], and the dispersion and losses can be included without violating canonical commutation rules. In this approach the photon momentum is not well-defined, in contrast to the mode decomposition technique. However, while certain restrictions on momenta can be important for some phenomena where atomic recoil should be taken into account, in the noisy solid state environment this is of less significance.

We consider a subwavelength light emitting system such as an atom or a quantum dot having a discrete set of levels $\{|n\rangle\}$ and placed at the point $\vec{r}_A$ in the photonic environment characterised by the (dispersive and lossy) dielectric function $\varepsilon(\vec{r},\omega)$ (e.g. Fig. 1). We restrict ourselves to the dipole transitions, characterized by the matrix elements $d_{n\to m}$, and apply a rotating wave approximation. The total Hamiltonian of the system is given by [20]:

$$\mathrm{H} = \mathrm{H}_0 + \mathrm{H}_{field} + \mathrm{H}_{int}$$
$$\mathrm{H}_0 = \sum_n \hbar \omega_n |n\rangle\langle n|$$
$$\mathrm{H}_{field} = \int d^3\vec{r} \int_0^\infty d\omega \cdot \hbar\omega f^\dagger(\vec{r},\omega) f(\vec{r},\omega)$$
$$\mathrm{H}_{int} = \int_0^\infty d\omega \cdot \int d^3\vec{r} \cdot V(\vec{r},\omega) \quad (1)$$
$$V(\vec{r},\omega) = -i\sqrt{\frac{\hbar}{\pi\varepsilon_0}} \sum_{m,n} \int_0^\infty d\omega \cdot \left(\frac{\omega}{c}\right)^2 \int d^3\vec{r} \cdot \sqrt{\mathrm{Im}\,\varepsilon(\vec{r},\omega)} \cdot d_{n\to m} G(\vec{r}_A,\vec{r},\omega) f(\vec{r},\omega) \sigma_{mn} + \mathrm{J}$$

where $\mathrm{H}_0$, $\mathrm{H}_{field}$, $\mathrm{H}_{int}$ are the Hamiltonians of the unperturbed atomic system, free electromagnetic field, and light-matter interaction respectively. $f^\dagger(\vec{r},\omega), f(\vec{r},\omega)$ are local creation and annihilation field operators dependent on position in space $\vec{r}$ and frequency $\omega$ and $\vec{G}(\vec{r}_A,\vec{r},\omega)$ is the classical electromagnetic Green's tensor defined by:

$$\left[\nabla\times\nabla\times - \left(\frac{\omega}{c}\right)^2 \varepsilon(\vec{r},\omega)\right]\vec{G}(\vec{r},\vec{r}_A) = \delta(\vec{r}-\vec{r}_A), \qquad (2)$$

where $\vec{r}_A$ is a source position and $\vec{r}$ is an arbitrary point in space. $\sigma_{mn} = |m\rangle\langle n|$, and the rest of the symbols are the widely used physical constants.

The spontaneous one-photon emission rate ($R_{SOPE}$) in the weak coupling regime and within the framework of the local operators' formalism is given by [20]:

$$R_{SOPE} = R_0 \frac{\vec{d}_{21}^{*}\vec{P}_1\vec{d}_{12}}{|d_{21}|^2}, \quad \vec{P}_1(\vec{r}_A,\omega_{if}) = \frac{3c}{2\omega_{if}}\operatorname{Im}\vec{G}(\vec{r}_A,\vec{r}_A,\omega_{if}) \qquad (3)$$

where $\omega_{if}$ is the frequency of the basic transition and $R_0$ is the vacuum decay rate $4\omega_{if}^3|d_{21}|^2/3\hbar c^3 \varepsilon_0$. Here, $P_1$ is the tensor of Purcell factors, determined by the imaginary part of the Green function. In the case of a single mode cavity with quality factor $Q$, modal volume $V$ and wavelength $\lambda$, $P_1$ is proportional to the well-known expression $Q \cdot \lambda^3 / V$. However, a crucial assumption of negligible losses in the material components is required and in particular, this simplified approach was used to fit the experimental data of [18] making it unsuitable for our purposes.

The STPE rate for the system described by Eq. (1) is calculated by the standard quantum mechanical second-order perturbation theory, utilizing the local quantization scheme. The STPE rate between initial and final states ($i \to f$) via complete set of intermediate states $\{|n\rangle\}$ is given by:

$$R_{STPE} = \frac{2\pi}{\hbar}\int_0^\infty d\omega \cdot \int d^3r \cdot \left|\sum_n \left(\frac{\langle f|V(\vec{r},\omega)|n\rangle\langle n|V(\vec{r},\omega)|f\rangle}{\omega_{in}-\omega+i\Gamma} + \frac{1}{\omega-\omega_{nf}+i\Gamma}\right)\right|^2 \delta(E_i - E_f - \hbar\omega) \qquad (4.0)$$

Using the Green function identities [20], one can express the result via the Purcell factor, Eq. (3),

$$R_{STPE} = \int_0^{\omega_{if}} d\omega \text{Tr}\left[\vec{\vec{P}}_1(\omega)\vec{\vec{P}}_1(\omega_{if}-\omega)\right] U(\omega),$$

$$U(\omega) = \frac{4\pi \omega_{if}^3 (\omega_{if}-\omega)^3}{9\hbar c^2 \varepsilon_0^2} \left| \sum_n d_{i\to n} d_{n\to f} \left( \frac{1}{\omega_{in}-\omega+i\Gamma} + \frac{1}{\omega-\omega_{nf}+i\Gamma} \right) \right|^2$$

(4.1)

Equation (4.1) has a transparent physical meaning: the STPE rate is given by the convolution of two one-photon Purcell factors, with the frequency-dependent factor sensitive to the relative energy positions of the states $|i\rangle, |f\rangle$ and $\{|n\rangle\}$. The difference between one- and two-photon Purcell effects is summarized in the schematic drawing on Fig. 2. While the one-photon emission is enhanced only at the resonance of the structure, STPE, being naturally wideband, is enhanced twice – at the resonant frequency ($\omega_{res}$) of a structure and at the complementary one ($\omega_0 - \omega_{res}$), emphasizing the 'paired' nature of this energy-time entangled emission process.

The line broadening $\Gamma$ of the emitter in Eq. 4 is associated with both radiative and nonradiative decay channels (including the interactions with an environment) that are taken account of phenomenologically by replacing discrete atomic levels by spectrally broad ones. The specific spectral shape ($\Gamma(\omega)$) may be calculated for each physical system, e.g. it has a Lorentzian shape for radiative broadening in a collisionless dilute gas, but will be more complicated for other systems, like solid state emitters [21]. The modified version of Eq. 4 for the spectral density, including these effects is:

$$U(\omega) = \int d\omega U(\omega, \Gamma) \Gamma(\omega_0 - \omega) \quad (5)$$

It is also important to assess the role of the losses in the medium, determined by the imaginary part of the dielectric constant. Only some of the emitted photons contributing to the true radiative decay may be observed in the far field, while the rest are absorbed by the lossy medium [22]. The radiative part of the Purcell factor may be then singled out as follows:

$$\vec{\vec{P}}_{1,RAD}(\vec{r}_A, \omega_{if}) = \vec{\vec{P}}_1(\vec{r}_A, \omega_{if}) - \frac{3\omega_{if}}{2c} \int d^3\vec{r} \cdot \mathrm{Im}\,\varepsilon(\vec{r}) \vec{\vec{G}}^*(\vec{r}_A, \vec{r}, \omega_{if}) \vec{\vec{G}}(\vec{r}, \vec{r}_A, \omega_{if}), \quad (6)$$

where the second term corresponds to the Joule losses. Substituting Purcell factor (6) in Eq. (4), we obtain the 'true' radiative contribution to the STPE decay rate. Eqs. 3, 7 and Fig. 2 provide the general concept for two-photon emission engineering. By properly tailoring the nanostructured environment by utilizing plasmonic, dielectric and Bragg resonances one can obtain a system with a quite general spectral response.

In the following, for the sake of simplicity, we will consider a specific case of a 3-level atom near a nanosphere (Fig.1). We consider a single intermediate state with the energy above both initial and final states (Fig. 1 inset) and neglect line broadening. To examine the effect of local field enhancement we have analyzed the case when the emitter is placed in a $SiO_2$ matrix in the proximity of both silicon and silver spheres of the same subwavelength dimensions. Metallic nanoparticles with negative permittivity may support collective oscillations of surface charges in the optical and infrared spectral range termed as localised surface plasmon resonances [23]. Such plasmonic particles can concentrate the optical field beyond the classical diffraction limit [24], while the resonant response of the dielectric particles with positive permittivity (like Si) may only result from wave interference, and hence is not possible at subwavelength dimensions. For the chosen geometry the electrodynamical Green function can be readily found as series over spherical harmonics. The dielectric function of Si and the Drude parameters of Ag were taken from widely available experimental data, while the value $\varepsilon = 2$ was chosen for the permittivity of $SiO_2$.

Fig. 3 presents the spontaneous one- and two -photon for a source located at different distances $d$ from the surface of the Ag particle. Panel (a) shows the calculated energy dependence of the one-photon Purcell factor (Eq. 3) along the line between emitter and nanoparticle centre. The peak locations correspond to the presence of plasmonic resonances, marked by $j = 1,..$ (dipole and higher). The low-energy maximum of the Purcell factor (thick lines on panel (a)) has a nonradiative origin and is related to the Joule heating of the metal. This maximum vanishes for the radiative part of the Purcell factor (6), shown by thin lines.

Panel (b) presents the normalized spectral dependence of STPE. These graphs possess characteristic mirror symmetry with respect to the central energy $\hbar \omega_{if} / 2$. The blue dashed curve in Fig. 3b corresponds to the STPE spectrum for an infinite distance $d$ between the atom and the nanoparticle, i.e. the scenario of an emitter embedded in an empty silica matrix. In this case the STPE spectrum is rather smooth and depends only on the 'free space like' photon density of states ($\sim \omega^2$). The highest emission probability corresponds to the degenerate case (equal energies of the emitted photons), so that the product of their densities of states is at a maximum (Eq. (4)). Fig. 3 shows a growth in the emission rates when the distance between the atom and the sphere surface becomes smaller than the particle radius. In the case of the Ag nanoparticle the overall integrated enhancement of STPE at $d=10$ nm is about 140-fold.

The STPE spectrum exhibits a rich multi-peak structure due to the plasmon resonances. This resonant structure can be most easily understood when compared with that in the one-photon spectra (Fig 3a). The two central maxima in Fig. 3b, marked by the arrows $j=1$, correspond to the electric dipole resonance, capable of emission enhancement at its eigen frequency ($\omega_d$) and complementary frequency ($\omega_{if}-\omega_d$). The bordering narrow peaks are related to higher order multipole resonances. One can see from Fig. 3b that the high order resonances become more pronounced when the atom approaches the boundary since multipole modes have a higher field localisation in close proximity to the particle. The transition energy $\hbar\omega_{if}$ in Fig. 3 was chosen to be slightly above the dipole and quadrupole plasmonic resonances of the nanoparticle, lying in the optical spectral range (Fig 3a). As a result, the counterparts of these plasmonic resonances manifest themselves in the STPE in the IR and THz ranges. This possibility to "convert" structure resonances from high to low frequencies (and vice versa) along with recent advances in the engineering of the spectral response of plasmonic particles [25,26] reveals the potential ability to design STPE spectra for specific applications [27].

The thin curves on Fig. 3b show the radiative contribution to the STPE rate, where Joule losses are substracted. A comparison between thick and thin lines reveals the significant role of the losses for low- and high- energy regions of the STPE spectrum which is explained by the efficient absorption of low-energy photons in metal (Fig. 3a). Nevertheless, the overall radiative STPE rate, as well the rate of emission of THz photons, is still enhanced when compared to that in bulk $SiO_2$ matrix.

It is also instructive to analyze the dependence of STPE on the spectral position of the plasmonic resonance with respect to the emission frequency (Fig. 4). The spectra, calculated for both Ag and Si particles at different transition energies $\hbar\omega_{if}$ with fixed energy of the intermediate state $\hbar\omega_{ni} = 0.3\,\text{eV}$, are presented on Fig 4. All of the graphs are normalized to the overall emission rate in the empty silica matrix. For the Si nanoparticle, the spectral dependencies are relatively smooth and structureless; the enhancement results from the density of states near the high-index dielectric boundary (Fig. 4b). The overall enhancement weakly depends on the transition energy. On the other hand, the plasmonic particle allows further improvement of the emission rate by a factor of up to 320. The origin of this effect is the spectral overlap between the two dipole resonances at $\omega_d$ and $\omega_{if} - \omega_d$. This may be further increased by placing the emitter even closer to the interface. Fig. 5 shows the radiative contribution to the STPE emission rate calculated for the parameters of Fig. 4. By comparing Fig. 5 and Fig. 4, one can see that despite a considerable suppression of the rate due to the losses, the effect of plasmonic resonances remains significant, giving rise to a complex STPE spectrum.

These examples demonstrate that the simplest spherical nanoparticle can dramatically modify the characteristic STPE properties of an emitter. A proper design of multiresonant plasmonic structures with variable and tunable quality factors may, in principle, provide any spectral response on demand. Moreover, in the STPE process, the emitted photons are always time-energy entangled and thus, such a multiresonant plasmonic environment engenders the possibility to achieve multiparticle entanglement.

In conclusion, we have developed a rigorous analytical approach to evaluate spontaneous two-photon emission rates in a structured metallo-dielectric photonic environment. The local field operators enable the formulation of this purely quantum problem in terms of classical electrodynamics by expressing the two-photon emission spectrum via the one-photon Purcell factors. As an example, the influence of localised surface plasmon resonances on the fundamental multi-photon processes was demonstrated by the full analytical calculation of emission rates in the vicinity of nanoscale particles. The unique nature of localised surface plasmons to confine light on subwavelength scales leads to at least 2-3 orders of magnitude enhancement of the emission together with on demand spectral reshaping in a broad spectral range.

Acknowledgements

This work was supported in part by the EPSRC (UK) and Australian Research Council (Australia). A. Poddubny acknowledges support from RFBR and the European projects POLAPHEN and Spin-Optronics. P. Ginzburg acknowledges the Royal Society for Newton International Fellowship and Yad Hanadiv for Rothschild Fellowship. A. Poddubny also acknowledges fruitful discussions with M.M. Glazov.

Figure captions:

**Figure 1** (color online) Two-photon dipole emitter in the proximity of nanosphere of radius $R$ ($d$ is the distance from the interface). (Inset) Level structure of the emitter: initial and final states - bold lines, the intermediate state - dashed line.

**Figure 2** Shows the difference between one- and two-photon Purcell factors. One-photon emission from equally distributed in frequency sources (first line) is enhanced at the structure resonances. Wideband STPE (second line) is enhanced twice – at the resonant frequency ($\omega_{res}$) of a structure and at the complementary ($\omega_0 - \omega_{res}$). Different scenarios are marked by set of red and blue arrows, staying for different resonances of the photonic structure.

**Figure 3** (color online) (a) shows the environmental contributions to the emission rate in terms of the Purcell factor for a dipole oriented perpendicular to an Ag sphere at various distances from its surface. The thick and thin curves correspond to the full Purcell factor, Eq. (3), and its radiative part, Eq. 6, respectively. (b) The corresponding spectral densities of STPE normalized to the overall emission rate in an empty silica matrix (i.e. the area under blue-dashed line is unity). Numbers near the curves show the overall STPE enhancement with respect to the bulk $SiO_2$ matrix. Thick and thin curves correspond to the total STPE rate and its radiative part. Vertical arrows indicate the energies of the electric multipole modes with different total angular momentum values $j$. All curves are calculated for different distances $d$ between the emitter in $SiO_2$ matrix and the nanoparticle surface.

**Figure 4** (color online) Normalised STPE spectra calculated for different transition energies $\hbar\omega_{if}$. (a) and (b) - Ag and Si nanoparticles in $SiO_2$ matrix, respectively. The photon energy on the *x*-axes is normalized to transition energy. Calculation parameters are the same as in Fig. 2b.

**Figure 5** (color online) The radiative part of the normalized STPE spectra calculated for different transition energies $\hbar\omega_{if}$ with Joule losses removed. Notation and calculation parameters are the same as in Fig. 4.

Figure 1.

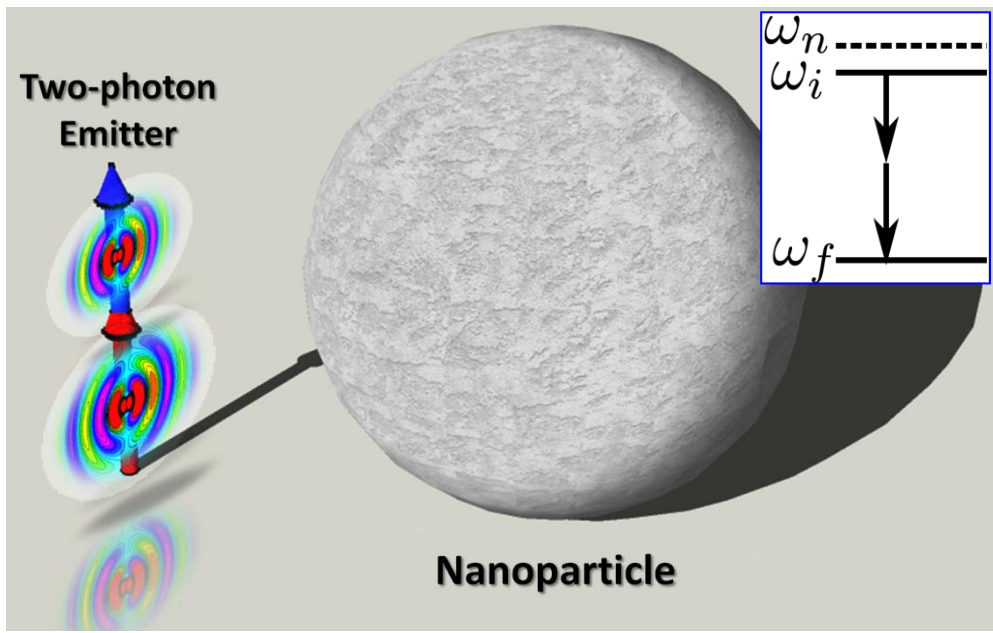

Figure 2.

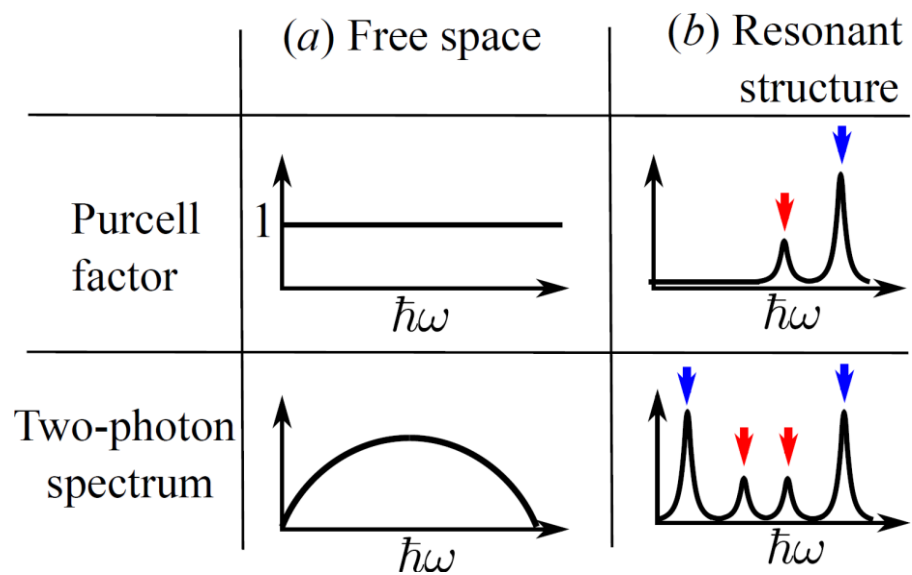

Figure 3.

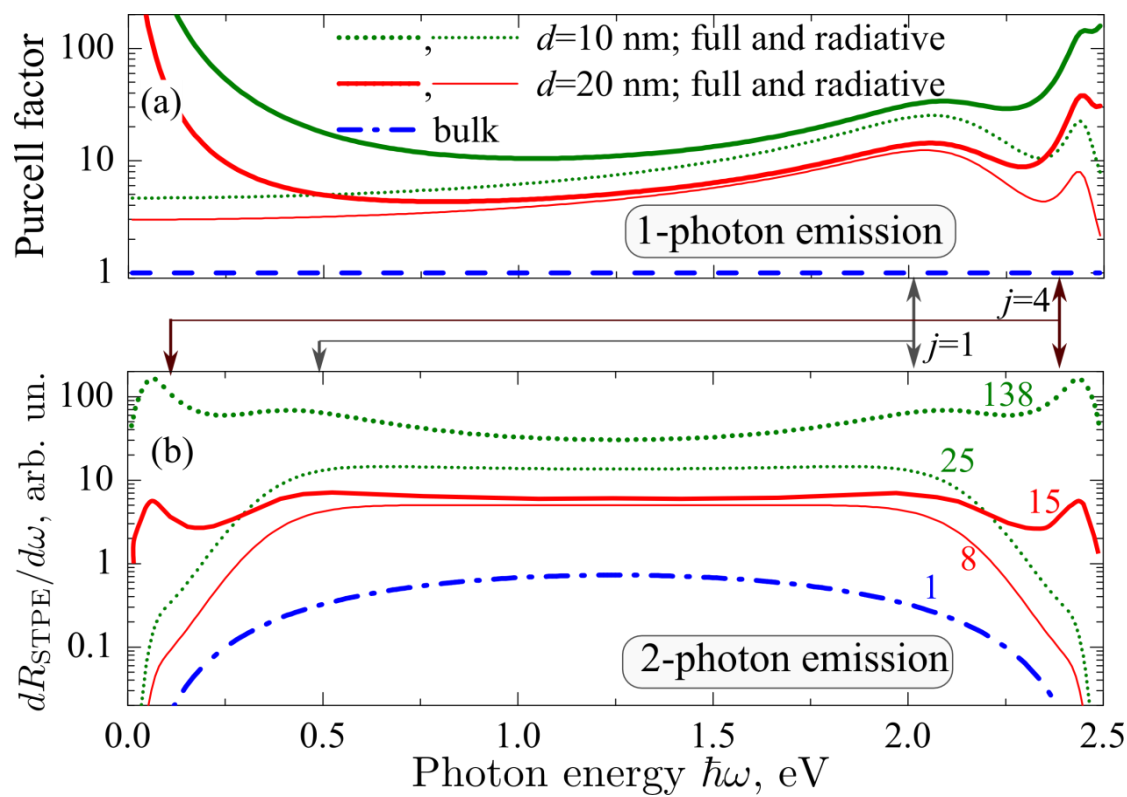

Figure 4.

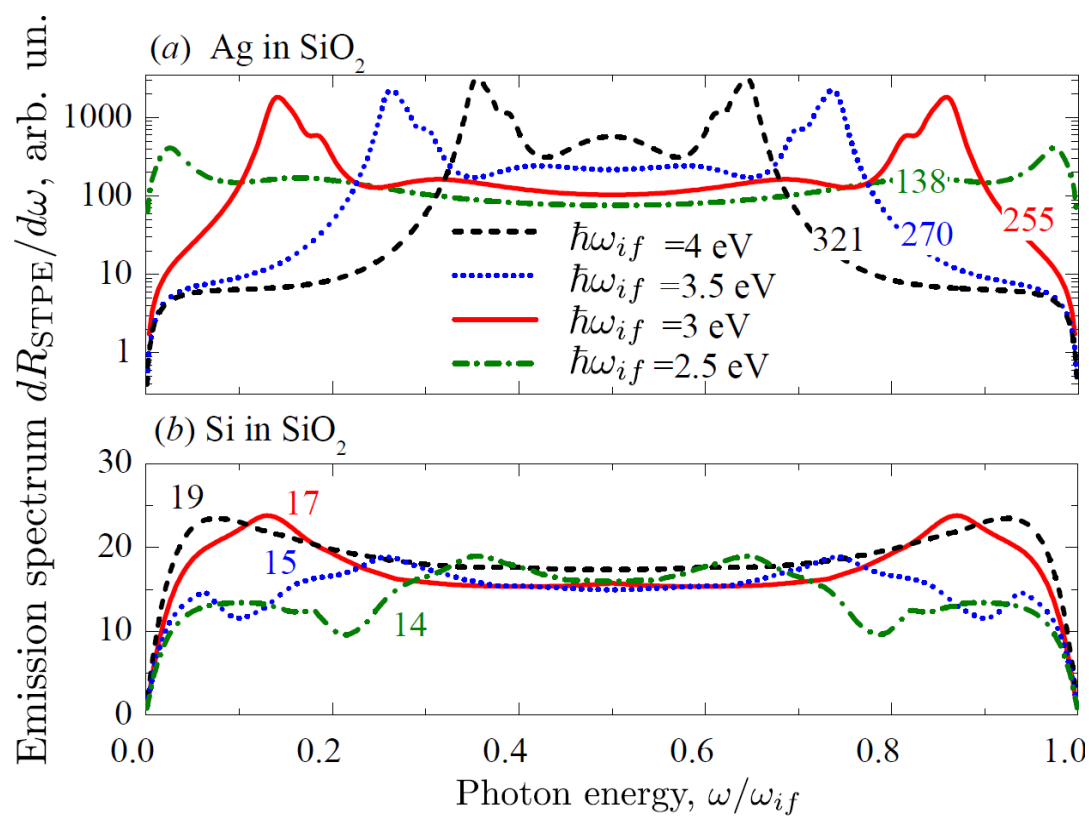

Figure 5.

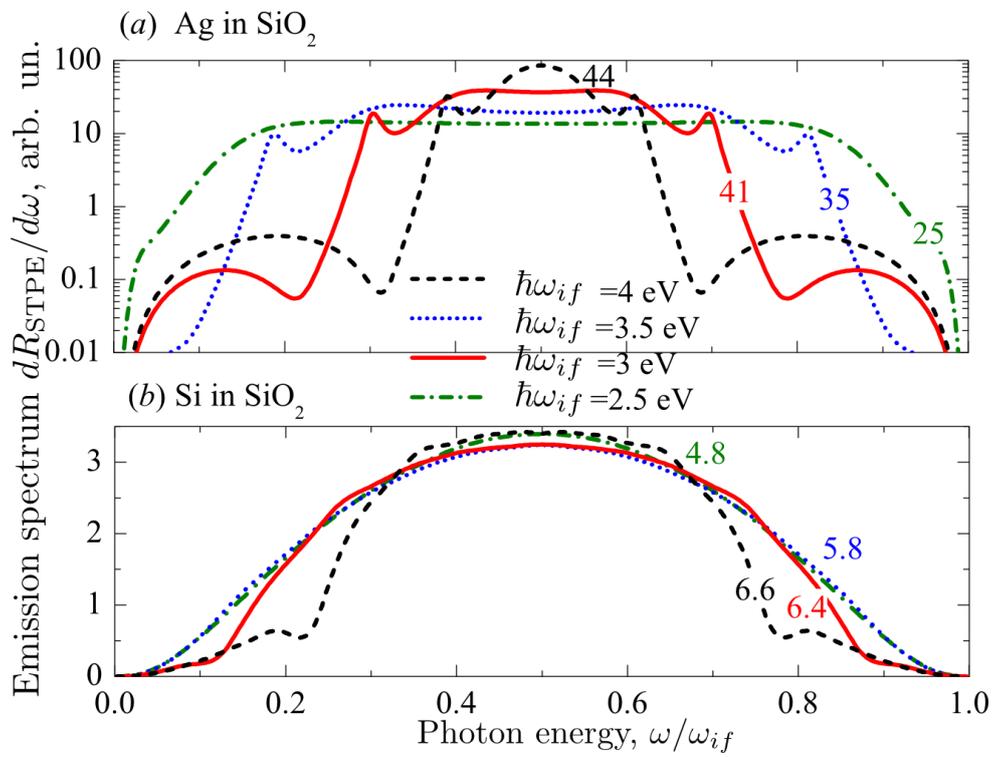

References:

1. P. G. Kwiat, K. Mattle, H. Weintfurter, A. Zeilinger, A. V. Sergienko, and Y. H. Shih "New High Intensity Source of Polarization-Entangled Photon Pairs": Phys. Rev. Lett., **75**, 4337 (1995).

2. A. Nevet, A. Hayat, and M. Orenstein, "Measurement of optical two-photon gain in electrically pumped AlGaAs at room temperature", Phys. Rev. Lett. **104**, 207404 (2010).

3. D. J. Gauthier, Q. Wu, S. E. Morin, and T. W. Mossberg, "Realization of a continuous-wave, two-photon optical laser", Phys. Rev. Lett. **68**, 464 (1992).

4. H. P. Yuen, "Two-photon coherent states of the radiation field", Phys. Rev. A **13**, 2226 (1976).

5. M. O. Scully, K. Wodkiewicz, M. S. Zubairy, J. Bergou, N. Lu, and J. Meyer ter Vehn, "Two-photon correlated–spontaneous-emission laser: Quantum noise quenching and squeezing", Phys. Rev. Lett. **60**, 1832 (1988).

6. F. Grosshans and P. Grangier, "Continuous Variable Quantum Cryptography Using Coherent States", Phys. Rev. Lett. **88**, 057902 (2002).

7. M. B. Nasr, D. P. Goode, N. Nguyen, G. Rong, L. Yang, B. M. Reinhard, B. E. A. Saleh, and M. C. Teich, "Quantum Optical Coherence Tomography of a Biological Sample," Opt. Commun. **282**, 1154 (2009).

8. A. Hayat, P. Ginzburg, M. Orenstein, "Observation of Two-Photon Emission from Semiconductors", Nature Photon. **2**, 238 (2008)

9. H. M. van Driel, "Semiconductor optics: On the path to entanglement", Nature Photon. **2**, 212 (2008).

10. A. Hayat, P. Ginzburg, and M. Orenstein, "High-Rate Entanglement Source via Two-Photon Emission from Semiconductor Quantum Wells", Phys. Rev. B **76**, 035339 (2007).

11. E. M. Purcell, "Spontaneous Emission Probabilities at Radio Frequencies", Phys. Rev. **69**, 681 (1946).

12. P. Lodahl, A. F. van Driel, I. S. Nikolaev, A. Irman, K. Overgaag, D. Vanmaekelbergh, and W. L. Vos, "Controlling the dynamics of spontaneous emission from quantum dots by photonic crystals", Nature **430**, 654 (2004).

13. L. A. Blanco and F. J. García de Abajo, "Spontaneous light emission in complex nanostructures", Phys. Rev. B **69**, 205414 (2004).

14. J. B. Khurgin, G. Sun, and R. A. Soref, "Enhancement of luminescence efficiency using surface plasmon polaritons: figures of merit", J. Opt. Soc. Am. B **24**, 1968 (2007).

15. Z. Jacob, J. Kim, G. V. Naik, A. Boltasseva, E. E. Narimanov, and V. M. Shalaev, "Engineering photonic density of states using metamaterials", Appl. Phys. B **100**, 215 (2010).

16. K. Tanaka, E. Plum, J. Y. Ou, T. Uchino, and N. I. Zheludev, "Multifold Enhancement of Quantum Dot Luminescence in Plasmonic Metamaterials", Phys. Rev. Lett. **105**, 227403 (2010).

17. A. N. Poddubny, P. A. Belov, and Y. S. Kivshar, "Spontaneous radiation of a finite-size dipole emitter in hyperbolic media", Phys. Rev. A **84**, 023807 (2011).

18. A. Nevet, N. Berkovitch, A. Hayat, P. Ginzburg, S. Ginzach, O. Sorias, and M. Orenstein, "Plasmonic nano-antennas for broadband enhancement of two-photon emission from semiconductors", Nano Lett. **10**, 1848. (2010).

19. N. A. R. Bhat, and J. E. Sipe, "Hamiltonian treatment of the electromagnetic field in dispersive and absorptive structured media", Phys. Rev. A **73**, 063808 (2006).

20. W. Vogel and D. Welsch, *Quantum Optics* (Wiley-VCH, 2006).

21. A. Hayat, P. Ginzburg, and M. Orenstein, "Measurement and model of the infrared two-photon emission spectrum of GaAs", Phys. Rev. Lett. **103**, 023601 (2009).

22. M. M. Glazov, E. L. Ivchenko, A. N. Poddubny, and G. Khitrova, "Purcell Factor in small metallic cavities", Phys. Solid State **53**, 1753 (2011).

23. S. A. Maier, *Plasmonics: Fundamentals and Applications* (Springer, New York, 2007).

24. D. K. Gramotnev and S. I. Bozhevolnyi, "Plasmonics beyond the diffraction limit", Nat. Photonics **4**, 83 (2010).

25. E. Prodan, C. Radloff, N. J. Halas, and P. Nordlander, "A hybridization model for the plasmon response of complex nanostructures", Science **302**, 419 (2003).

26. P. Ginzburg, N. Berkovitch, A. Nevet, I. Shor, and M. Orenstein, "Resonances On-Demand for Plasmonic Nano-Particles", Nano Lett. **11**, 2329 (2011).

27. A. V. Kavokin, I. A. Shelykh, T. Taylor, and M. M. Glazov, "Vertical Cavity Surface Emitting Terahertz Laser", Phys. Rev. Lett. **108**, 197401 (2012).